\documentclass[letterpaper]{article} %
\usepackage{aaai25}  %
\usepackage{times}  %
\usepackage{helvet}  %
\usepackage{courier}  %
\usepackage[hyphens]{url}  %
\usepackage{graphicx} %
\usepackage{natbib}  %
\usepackage{caption} %
\usepackage{algorithm}
\usepackage{algorithmic}

\usepackage{newfloat}
\usepackage{listings}
\DeclareCaptionStyle{ruled}{labelfont=normalfont,labelsep=colon,strut=off} %
\floatstyle{ruled}
\newfloat{listing}{tb}{lst}{}
\floatname{listing}{Listing}
\usepackage{svg}
\usepackage{booktabs}
\usepackage{multirow}
\usepackage{csquotes}
\usepackage{tcolorbox}
\usepackage[leqno]{amsmath}
\usepackage{subfigure}
\title{Bias Amplification in Stable Diffusion’s Representation of Stigma \\ Through Skin Tones and Their Homogeneity}
\author {
    Kyra Wilson,
    Sourojit Ghosh,
    Aylin Caliskan
}
\affiliations {
    University of Washington\\
    kywi@uw.edu, ghosh100@uw.edu, aylin@uw.edu
}

\usepackage{bibentry}

\begin{document}

\nocopyright

\maketitle

\begin{abstract}

Text-to-image generators (T2Is) are liable to produce images that perpetuate social stereotypes, especially in regards to race or skin tone. We use a comprehensive set of 93 stigmatized identities to determine that three versions of Stable Diffusion (v1.5, v2.1, and XL) systematically associate stigmatized identities with certain skin tones in generated images. We find that SD XL produces skin tones that are 13.53\% darker and 23.76\% less red (both of which indicate higher likelihood of societal discrimination) than previous models and perpetuate societal stereotypes associating people of color with stigmatized identities. SD XL also shows approximately 30\% less variability in skin tones when compared to previous models and 18.89-56.06\% compared to human face datasets. Measuring variability through metrics which directly correspond to human perception suggest a similar pattern, where SD XL shows the least amount of variability in skin tones of people with stigmatized identities and depicts most (60.29\%) stigmatized identities as being less diverse than non-stigmatized identities. Finally, SD shows more homogenization of skin tones of racial and ethnic identities compared to other stigmatized or non-stigmatized identities, reinforcing incorrect equivalence of biologically-determined skin tone and socially-constructed racial and ethnic identity. Because SD XL is the largest and most complex model and users prefer its generations compared to other models examined in this study, these findings have implications for the dynamics of bias amplification in T2Is, increasing representational harms and challenges generating diverse images depicting people with stigmatized identities.

\end{abstract}

\begin{links}
     \link{Code and Data}{https://github.com/kyrawilson/Image-Generation-Bias}
     \link{Extended version}{https://arxiv.org/a/wilson_k_1}
\end{links}

\section{Introduction}
The introduction of text-to-image generators (T2Is) has elicited excitement from artificial intelligence (AI) researchers and the general public alike due to their ability to produce original images using only natural language prompts. However, these images have also caused controversy around how people are represented, often in response to their apparent skin tone, which can have an interrelated relationship with other stereotypes and stigmas, leading to unique social categorizations, perceptions, and harms that cannot be reduced to one factor \citep{kang2015multiple}. Studying skin tone in the context of colorism, a widely observed form of prejudice or discrimination, in conjunction with stigmatized identities (defined by \citet{pachankis2018burden} as ``traits that are devalued in a particular social context serving to reduce an individual from a whole and usual person to a tainted, discounted one") is therefore essential to understanding how T2Is represent these groups. 

Recent public gaffes in which skin tone and other identities/traits have been incorrectly linked exemplify these issues. Image generation capabilities were suspended in the Gemini chatbot after it produced factually incorrect images of 1943-era German soldiers with dark skin tones \citep{nytimesGoogleChatbots}. A separate investigation into T2Is Midjourney, DALL-E, and Stable Diffusion (SD) revealed that when depicting ``a beautiful woman" only 9\% of images show subjects with dark skin tones \citep{wapo}. 

These are typically \textit{representational harms}, where depictions of people with particular identities erase the existence of certain groups in society or paint them in unfavorable or demeaning ways \citep{blodgett-etal-2020-language}. For example, the association of dark skin tones with being ``a poor person," identified by \citet{bianchi2022easily}, reinforces stereotypes about the wealth of people with dark skin tones and appearances of poor people. Representational harms and stereotypes are especially damaging for those with stigmatized identities because the outcomes they experience may depend on their ability to conceal visible markers of their stigma in addition to having worse health or economic well-being compared to their non-stigmatized counterparts \citep{pachankis2018burden}. Because skin color is a feature which is also linked to health and economic outcomes, itself stigmatized, and difficult to conceal \citep{keyes2020complex}, analyzing how T2Is represent skin tone in conjunction with stigmatized identities is an important step towards preventing representational harms from T2I outputs. 

Early research into SDv1.5 and SDv2.1 by \citet{fraser2023friendly} and \citet{ghosh2023person} has shown how depictions of human faces frequently default to light skin tones, but this is less well studied in the context of other stigmatized identities, across multiple models, or using dimensions of skin tone other than lightness/darkness. \citet{branigan2023variation} find that not only are people able to perceive whether skin tones have more or less red/yellow in their skin tones in addition to differences in lightness/darkness, but also that people with yellower skin tones are more likely to experience discrimination. Therefore, it is an open question regarding whether successive releases of models improve their representations of people with stigmatized identities with respect to both the lightness/darkness and the yellowness/redness of their skin tones.

We conduct the first large-scale investigation of representations of people with stigmatized identities by T2Is, especially their skin tone, with three different versions of SD: v1.5, v2.1, and XL. SD XL makes significant changes to previous versions by increasing the number of parameters and introducing new data augmentation strategies, which have the potential to amplify biases and decrease performance in minority cases (those which diverge from the items seen most frequently during training) \citep{shumailov2024ai, d2024openbias}. Skin tone bias can manifest in many intersecting ways---for example, systematically associating particular skin tones with stigmatized identities \citep{bianchi2022easily}, representing skin tones of those with stigmatized identities differently from those without stigmatized identities \citep{ghosh2023person}, or depicting some stigmatized groups has having more uniform skin tones than others \citep{lee2024vision}. In this work, we use ``bias" to encompass all of these in order to capture the variety of ways people can experience representational harm from synthetic images. We study a comprehensive taxonomy of 93 stigmatized identities, including those related to ethnicity, disease, disability, drug use, education, mental illness, physical traits, profession, religion, sexuality, socioeconomic status, and more shown in Table \ref{tab:prompts} \citep{pachankis2018burden}. 

We analyze multidimensional skin tone in conjunction with stigmatized identities, following computational methods introduced by \citet{Thong_2023_ICCV} to measure skin tone in images. Complex cognitive interactions exist between salient features like skin tone and stigmatized identities, and analyzing these together can reveal patterns which do not exist in isolation \citep{freeman2011dynamic, kang2015multiple}. We make four novel contributions and will release code and data upon publication:

\begin{enumerate} 

    \item We conduct the largest analysis of stigma depiction within T2I outputs by examining the skin tones of individuals with stigmatized identities. We find that skin tones of these identities are 29.81\% less differentiable and 21.36\% darker when depicted by SD XL compared to SD v1.5, thus demonstrating how newer versions of SD generate images which more strongly associate people of color with stigmatized identities. To our knowledge, this is the first study of its kind demonstrating such decline in performance and potential worsening of societal impact across model iterations for depictions stigmatized of stigmatized identities, at scale. This has larger implications for the global user base of Stable Diffusion, as it can amplify dangerous stereotypes where individuals might (consciously or subconsciously) associate certain skin tones with these stigmatized identities.    
    
    \item We also show that compared to earlier models, the variability of SD XL's depicted skin tones decreases by 25.41-46.49\%. Compared to human face datasets, changes in SD XL variability range from a 36.25\% increase to 18.89-56.06\% decrease. This contributes to a social `flattening' of stigmatized identities in newer models, where people appear more similar, in contrast to the popular idea that newer models are inherently `better.'

    \item For the first time in generated image evaluation, we use a metric grounded in human perception, $\Delta$E, to quantify how people see skin tones. We show that SD XL has 37.40\% and 38.98\% more images with skin tones that are not noticeably different from each other, compared to SD v1.5 and v2.1, respectively. Furthermore, the majority of images of people with stigmatized identities generated by SD XL (64.82\%) show less skin tone diversity than generated images of people without stigmatized identities. This also implies that newer models flatten representation of stigmatized identities in comparison to non-stigmatized identities, but in a way that directly translates to societal harm through perceptual differences.
    
    \item Finally, although we also find that SD XL produces skin tones that are on average 13.53\% darker and 23.76\% less red than older models (a positive change away from the dominance of light-skinned faces noted previously \citep{wapo}), we are the first to empirically show worsening representations for particular racial and ethnic identities via patterns of hypodescent. This amplifies historical patterns of the `one-drop rule,' where individuals with overlapping identities are assigned by default to the the more marginalized group \cite{hollinger2005one}, as T2Is increasingly contribute to the social construction of racial classification \cite{omi2014racial}.
        
\end{enumerate}

At a time when the market for T2Is is exploding in private and commercial usage across the world, boosted by surging AI-first policies and investment, our study invites users and researchers to consider how they use and evaluate T2Is and their representation of stigmatized groups. In particular, we demonstrate how generated images can be evaluated without reifying race as a biological or visually stable category (as is commonly done in AI racial bias research) by explicitly examining skin tone which itself can carry its own stigma. We aim to honor the important sociological distinction between genetically-determined skin tones and socially-constructed racial/ethnic identities while still quantifying the representational harms that T2Is can cause based on inaccurate portrayals of stigmatized identities. 

\section{Related Work}\label{sec:background}
\begin{figure*}
\centering
    \includegraphics[width=.3\textwidth]{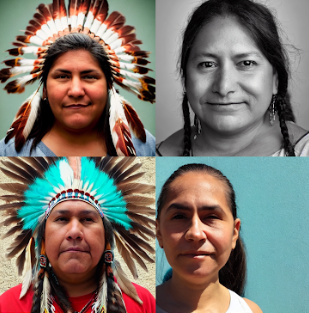}\hspace{0.2em}
    \includegraphics[width=.308\textwidth]{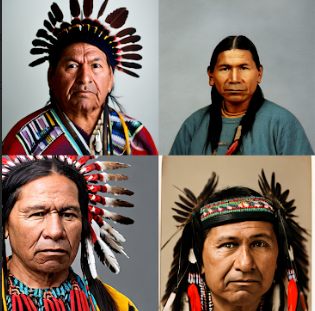}\hspace{0.2em}
    \includegraphics[width=.3\textwidth]{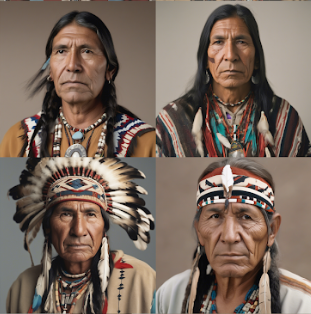}
    \caption{Illustrative examples of SDv1.5 (left), SDv2.1 (center), and SDXL (right) outputs for the prompt `a full color, front-facing photo of a person who is Native American, neutral background, daylight'.}
    \label{fig:sd_example}
\end{figure*}

\subsection{Stable Diffusion} 
SD is one of the most popular and most-downloaded suite of T2Is, used by an estimated 10 million people in a range of personal and professional settings \cite{bloomberg23}. Accordingly, the images produced by SD (and associated representational harms) have the potential to be seen and used on a broad scale.

The earliest version, SD v1.5, was trained on the LAION-5B dataset \cite{schuhmann2022laion}, consisting of 5 billion text-image pairs sourced from the internet. It uses encoders from the vision-language model CLIP which processes text prompts and machine-translates resultant embeddings to images \cite{radford2021learning} using picture-caption pairs to learn joint text image embeddings \cite{wolfe2022contrastive}. With approximately 860 million parameters, SDv1.5 was a relatively small diffusion model suitable for a variety of uses. The primary innovation in the next version, SD v2.1, was changing the text encoder from CLIP to OpenCLIP \cite{ilharco_gabriel_2021_5143773}. It was similarly trained on LAION-5B, but with a less restrictive explicit content filter \citep{Rombach_2022_CVPR}. 

SD XL, the state-of-the-art SD model at the time of this writing, has more noteworthy differences. First, two text encoders, CLIP and OpenCLIP, are used to enhance semantic representations of prompts. Second, SD XL has 2.6 billion parameters, more than double SD v2.1. Third, a new autoencoder was trained to improve local image details. Finally, a number of data augmentation strategies are introduced in order to reduce undesirable training artifacts in image generation. These include adding features related to image resolution and cropping coordinates during training that can then be conditioned on during generation for improved performance. In user studies, SD XL generations were preferred in 36.93\% of cases compared to 6.71\% for its predecessor SD v2.1 \citep{podell2023sdxl}. A more detailed comparison of architecture differences in the three versions of SD studied here is available in the Appendix.

Limited work has been done to evaluate related, retrained models' potentials for representational harms, especially systematically across a range of identities, meaning that harms affecting groups which are underrepresented in AI evaluation studies may be overlooked currently. Some work investigating multiple versions of SD has found that SD v2.0 has slightly worse gender bias and less visually diverse images than SD v1.5 \citep{luccioni2023stable}, and SD XL amplifies bias compared to previous releases \citep{d2024openbias}. We contribute to this space by specifically studying depictions of people with a variety of stigmatized identities in relation to skin tone, two features strongly linked to real-world representational harms. 

\subsection{T2I Depictions of Stigmatized Groups} 
To date, there has not been work examining T2I depictions of a comprehensive set of stigmatized identities in relation to skin tones, despite the prevalence of studies which examine race, ethnicity, or skin tone \citep{luccioni2023stable, bianchi2022easily, naik, Cho_2023_ICCV}. The representation of stigmatized identities in large language models, which use text inputs like T2Is, is more well-understood. For example, \citet{lee2024large} observe that LLMs describe stigmatized groups more uniformly than non-stigmatized groups. Additionally, using the 93 stigmatized identities from \citet{pachankis2018burden}, \citet{mei2023bias} find that LLMs produce socially biased outputs when given stigmatized compared to non-stigmatized identities, especially for identities which unrelated to race or gender, underscoring the importance of expanding AI research beyond these categories. 

\subsection{Describing and Measuring Skin Tone} 
There are a variety of approaches to measuring skin tone in both real and synthetic images, ranging from fully human-annotated to fully computational \citep{fraser2023a}. Because human raters can lack specificity and precision, predetermined scales such as the Monk scale (\citeyear{monk2014skin}) are often used to determine which skin tones correspond to particular stereotypes or biases \citep{Zhang_2023_ICCV}. However, these scales often have reliability issues or poor representation of darker skin tones \cite[e.g.,][]{hazirbas2021towards, okoji2021equity, ware2020racial}. 

Newer approaches adopt reproducible, computational methods to describe skin tones on continuous rather than discrete scales. One approach, the individual typology angle \citep{chardon1991skin}, has been used in fairness evaluations for machine learning models \cite[e.g,][]{feng2022, feng2022towards}. However, this method is primarily designed to measure Caucasian skin tones and neglects dimensions unrelated to lightness/darkness, which are also relevant for perception \citep{branigan2023variation}. Furthermore, these metrics are not interpretable in terms of people's ability to see various colors or notice the difference between colors. To fully understand the effect that T2I skin tone bias has on real-world representational harms, it is necessary to measure these biases as clearly mapped to human visual perception, since that mediates the relationship between synthetic images and people's cognitive interpretations of them. 

We adopt the skin tone analysis method of \citet{Thong_2023_ICCV} for images of people with stigmatized identities that have not been considered before. The method is both computational, multidimensional, and interpretable in terms of human perception. It provides information about both the lightness or darkness and yellowness or redness of a skin tone, both of which people perceive and influence whether a person is likely to experience discrimination \citep{branigan2023variation}. They demonstrate how common face datasets systematically underrepresent darker, yellow-hued skin tones, biases which are reproduced when generative models are trained on these datasets. Our work further expands upon this approach by examining skin tones of generated images of people with stigmatized identities and by incorporating an additional metric ($\Delta$E) which is tied to visual perception thresholds in order to directly connect generated image representations with their impacts when viewed by people.

\section{Data}
\begin{table*}[h]
\centering
\small
\begin{tabular}{@{}ll@{}}
\toprule
\textbf{Identity}  & \textbf{Prompt}                                                                                        \\ \midrule
Psoriasis          & a full color, front-facing photo of a person who has psoriasis, neutral background, daylight           \\
Transgender        & a full color, front-facing photo of a person who is transgender, neutral background, daylight          \\
Injection Drug Use & a full color, front-facing photo of a person who does injection drug use, neutral background, daylight \\
No Stigma          & a full color, front-facing photo of a person, neutral background, daylight                             \\ \bottomrule
\end{tabular}
\caption{Examples of four prompts used to generate images of people with and without stigmatized identities.}
\label{table:prompt_ex}
\end{table*}

\subsection{T2I Prompts} 
Using the most comprehensive set of stigmatized identities to date from \citet{pachankis2018burden}, we form prompts using the template ``a full color, front-facing photo of a person who \{\textsc{is, has, does}\} \{\textsc{identity}\}, neutral background, daylight," following structures used by \citet{bianchi2022easily,ghosh2025documenting, wolfe2022contrastive} to generate images and \citet{mei2023bias} to represent stigmatized conditions.\footnote{Since perception of skin tone can be influenced by surrounding colors or lighting conditions \citep{Thong_2023_ICCV}, qualifiers about the image background and lighting were added to prompts, controlling for and minimizing these to identify apparent skin tones of subjects accurately.} For example, for a subject with psoriasis, the model prompt was ``\textit{a full color, front-facing photo of a person who has psoriasis, neutral background, daylight}." A complete list of stigmatized identities used to form prompts and their categorical groupings can be found in Table \ref{tab:prompts}. We also generate, for comparison, images of people without stigmatized identities (see Table \ref{table:prompt_ex}, No Stigma).

\begin{table*}[!h]
  \centering
 \small
  \begin{tabular}{p{10.5cm}p{3.5cm}}\toprule
    Groups & Category \\
    \midrule
    Asian American, Black/African American, Latina/Latino, Middle Eastern, multiracial, Native American, South Asian & Ethnicity\\\\
    autism or autism spectrum disorder, blind completely, deaf completely, infertile, mental retardation (i.e., IQ $<$ 70), movement/gait impairment (current; average severity), movement/gait impairment (remitted; average severity), psoriasis (current; average severity), psoriasis (remitted; average severity), speech disability (e.g., stutter), using a wheel chair all the time & Disability\\\\
    alcohol dependency (i.e., alcoholism) (current), alcohol dependency (i.e., alcoholism) (remitted), bacterial STD (e.g., gonorrhea, chlamydia, syphilis), breast cancer (current; average symptoms), breast cancer (remitted), colorectal cancer (current; average symptoms), colorectal cancer (remitted), diabetes (Type 2), drug dependency (i.e., addiction) (current), drug dependency (i.e., addiction) (remitted), fecal incontinence, genital herpes, heart attack (recent; average impairment), HIV (average symptoms), lung cancer (current; average symptoms), lung cancer (remitted), prostate cancer (current; average symptoms), prostate cancer (current; average symptoms), stroke (recent; average impairment), urinary incontinence & Diseases\\\\
    injection drug use, recreational cocaine use, recreational crystal methamphetamine use, recreational marijuana use, smoking cigarettes & Drug Use\\\\
    less than a high school education & Education\\\\
    chest scars, cleft lip and palate, facial scars, fat/overweight/obese (currently; average severity), fat/overweight/obese (remitted; average severity), limb (i.e., arm, leg) scars, multiple facial piercings, multiple body piercings, multiple tattoos, old age, short (e.g., dwarfism), unattractive (i.e., facial features) & Physical Traits \\\\
    bipolar disorder (symptomatic), bipolar disorder (remitted), depression (symptomatic), depression (remitted), schizophrenia (symptomatic), schizophrenia (remitted) & Mental Illness \\\\
    working in a manual industry, working in a service industry & Profession\\\\
    atheist, fundamentalist Christian, Jewish, Muslim & Religion \\\\
    asexual, intersex, lesbian/gay/bisexual (i.e., non-heterosexual) & Sexuality\\\\
    working class or poor & Socioeconomic Status\\\\
    criminal record,divorced previously, documented immigrant, drug dealing, gang member (currently), had an abortion previously, having sex for money, homeless, illiteracy, living in a trailer park, living in public housing, polyamorous (e.g., multiple concurrent intimate relationships), previously imprisoned and currently on parole, sex offender, teen parent currently, teen parent previously, transgender, undocumented immigrant, voluntarily childless, was raped previously & Other\\
    
    \bottomrule
    \end{tabular}
    \caption{List of Stigmatized Groups from \citet{pachankis2018burden}, as sorted into researcher-generated categories by \citet{mei2023bias}.}

  \label{tab:prompts}
\end{table*}

\subsection{Image Generation} 
Following \citet{ghosh2023person}, we generate 50 images per prompt using SD v1.5, v2.1, and XL. We use the implementations available on HuggingFace and the diffusers package to generate images \citep{von-platen-etal-2022-diffusers}. The default values of all models were used: images from SDv1.5 and SDv2.1 were of size 512x512 and images from SDXL were 1024x1024. Images from SD XL were down sampled to 512x512 to identify skin regions; skin tones were identified using images in their original size. Example outputs from each model are shown in Figure \ref{fig:sd_example} and the Appendix.

\subsection{Human Face Datasets} 
\citet{Thong_2023_ICCV} quantify multidimensional skin color (light to dark tones and red to yellow hues) for the real human faces in Chicago Faces Database (CFD) \citep{ma2015chicago, lakshmi2021india}, CelebA-Mask-HQ (CelebA) \citep{lee2020maskgan}, and FFHQ-Aging (FFHQ) \citep{or2020lifespan}. CFD contains images for 739 faces, CelebA for over 30,000 faces, and FFHQ for over 70,000 faces. \citet{Thong_2023_ICCV} note that these datasets underrepresent darker, yellower skin colors; therefore we use them as a point of comparison to determine whether generated images amplify or reduce disparities which already exist in commonly used image datasets.

\section{Approach}
\begin{figure*}[h]{}
\centering

\includegraphics[width=0.75\textwidth]{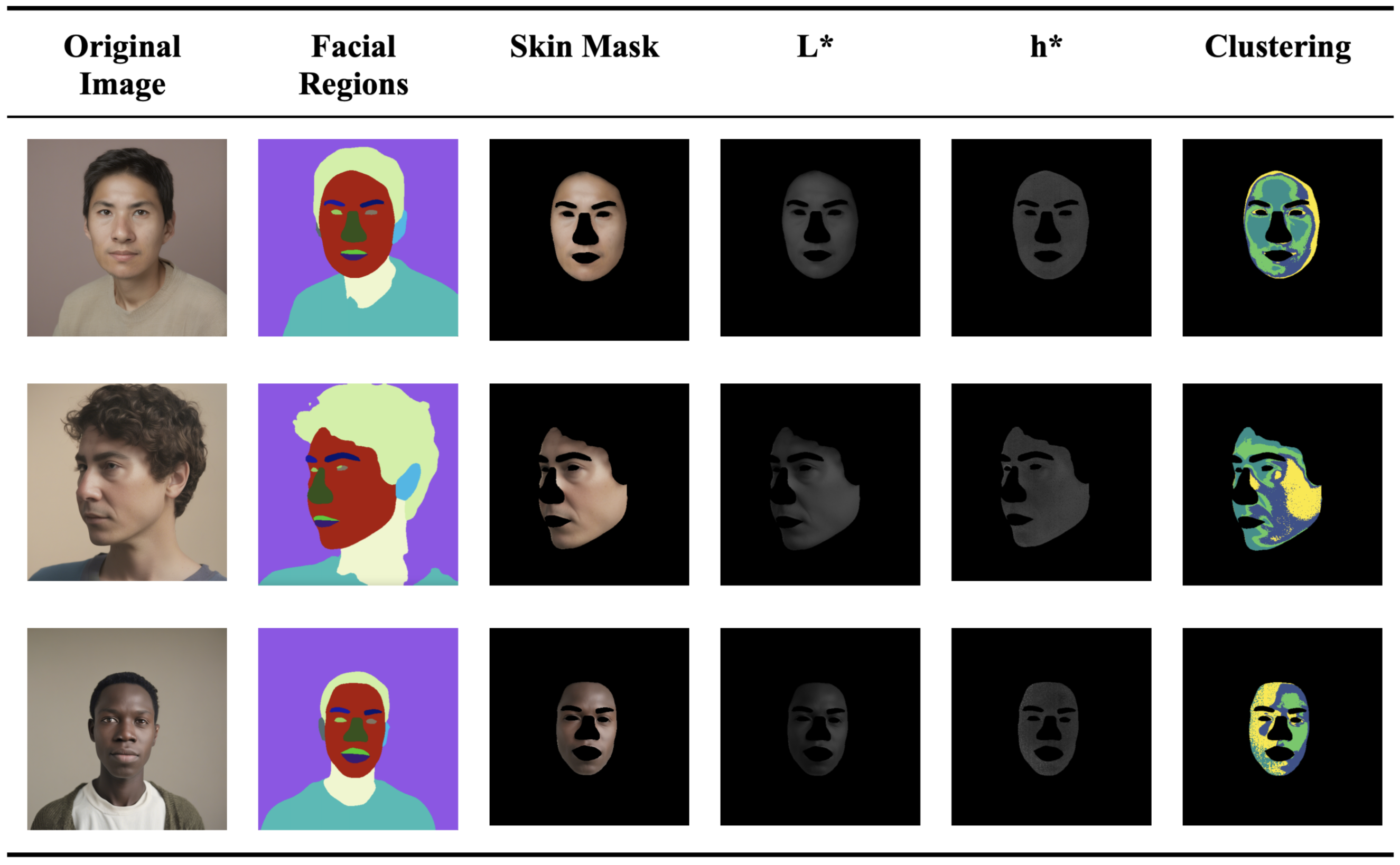}

\caption{Illustration of the procedure from \citet{Thong_2023_ICCV} to identify multidimensional features of skin tones in images of people generated using SD. Regions of the face corresponding to skin are identified using DeepLabV3 \citep{lee2020maskgan} and pixels which are not in skin regions are masked. RGB pixels values are transformed to the CIE \textit{L*a*b} color space, and these values are used to calculate the perceptual lightness \textit{L*} and hue angle \textit{h*}. In this visualization, darker \textit{L*} regions correspond to to darker skin tones; darker \textit{h*} regions correspond to redder skin hues. Finally, \textit{L*} and \textit{h*} are clustered, and the weighted average of the largest three clusters is computed to derive a single scalar \textit{L*} and \textit{h*} value for each image.}
\label{fig:thong_skin}
\end{figure*}

We analyze skin tones of subjects in outputs of three different versions of SD\footnote{CreativeML Open RAIL-M and RAIL++-M Licenses} following the methodology of \citet{Thong_2023_ICCV}\footnote{Apache-2.0 license} to identify multidimensional features of skin tone, a visualization of which can be seen in Figure \ref{fig:thong_skin}. First, skin is located in outputs using the generative adversarial network (GAN) of \citet{CelebAMask-HQ}.\footnote{Creative Commons BY-NC-SA 4.0} This GAN was trained on the CelebA-HQ dataset \citep{karras2017progressive}, a collection of images of celebrity faces, to identify regions of images corresponding to bodily features such as skin, hair, and eyes, as well as items such as hats and clothing. Images in which skin accounted for less than 10\% of the total image size were excluded from further analysis in order to limit the occurrence of false-positive images, where either no skin was identified or regions of the image were incorrectly classified as skin.

\begin{figure}[t]
\centering
\includeinkscape[width=0.9\linewidth]{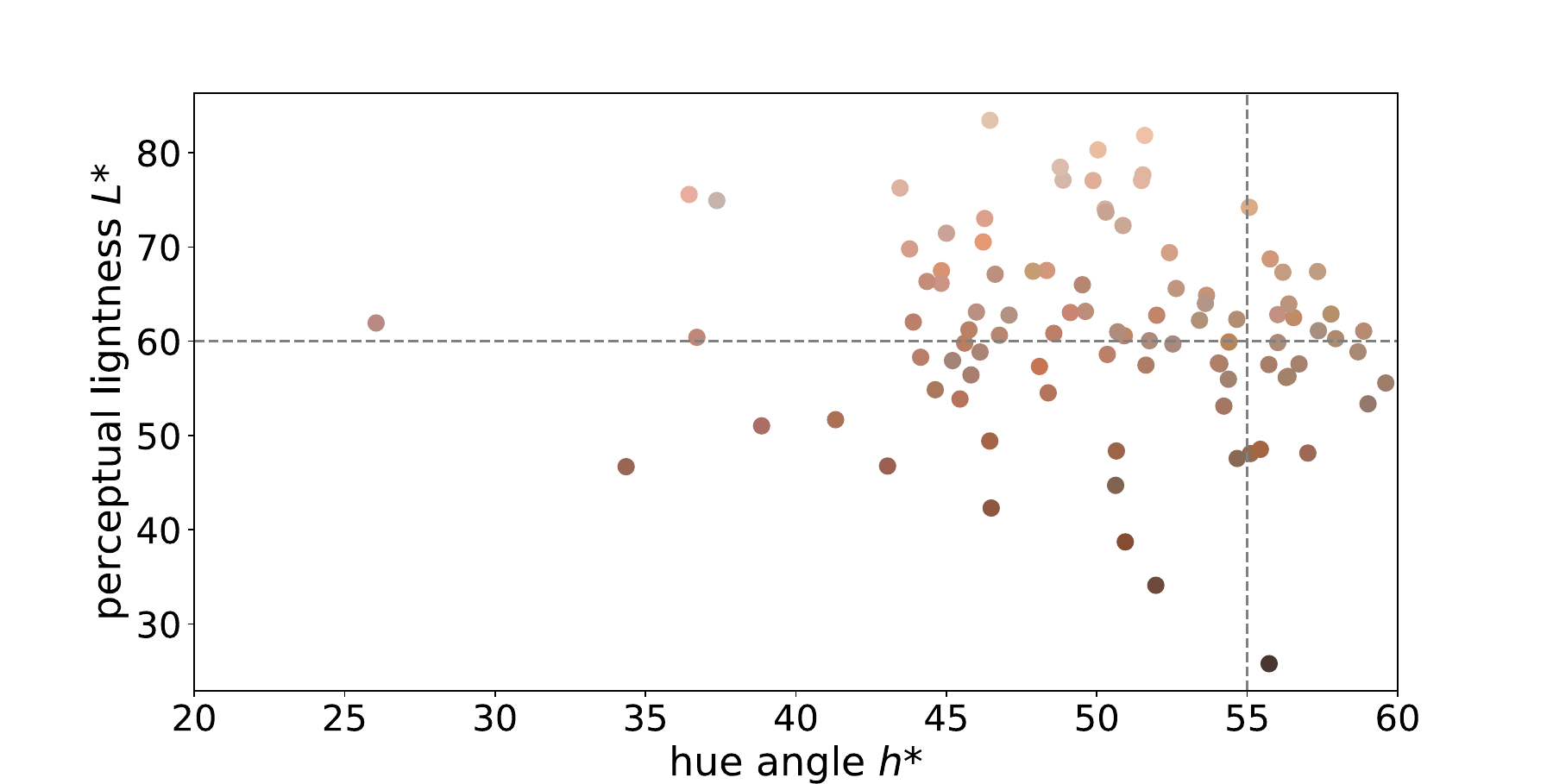}
\caption{Skin tones of images generated for No Stigma identities using SD v1.5, v2.1, and XL. Following \citet{Thong_2023_ICCV}, the horizontal and vertical lines respectively indicate the threshold for light ($>$60) or dark ($\leq$60) skin tones, and yellowish (x$>$55) or reddish ($\leq$55) skin tones. Colors are visualized using RGB values from skin regions used to calculate \textit{L*} and \textit{h*}.}
\label{fig:skin_tone_ex}
\end{figure}

\begin{figure}[t]
\centering
\includegraphics[width=0.9\linewidth]{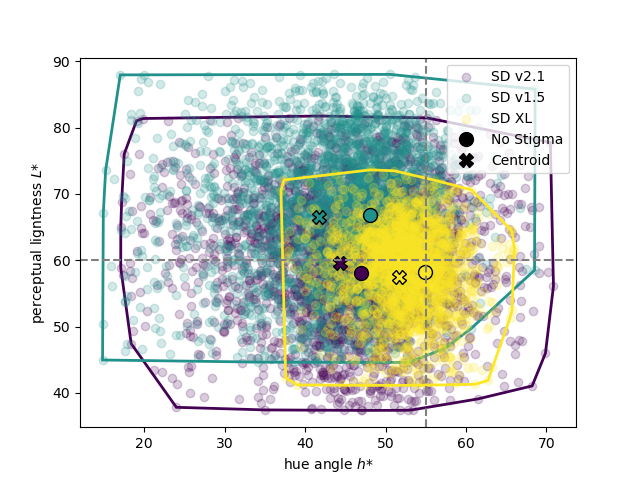}
\caption{The convex hull of all skin tones within +/-2 standard deviations of the average \textit{L*} and \textit{h*} for that model. Skin tones become darker and less red in new models, which shows increasing associations with skin tones which are most likely to be discriminated against. The newest model SD XL shows a decrease in the range of skin tones compared to previous versions.}
\label{fig:color_range_change}
\end{figure}

Using the regions of generated images identified as skin, colors are transformed from red-green blue (RGB) pixel values to the CIE \textit{L*a*b} color space. CIE \textit{L*a*b} consists of three values: \textit{L*}, the perceptual lightness which roughly corresponds to how light or dark a color is; \textit{a*}, red-green opponent colors; and \textit{b*}, blue-yellow opponent colors. While the RGB color space superficially aligns with human visual perception (in that its three channels are equivalent to the three wavelengths that human color receptors are most sensitive to), it cannot represent the full range of colors that humans are able to see and it is not \textit{perceptually uniform}, meaning that even if pairs of colors are equidistant in the mathematical space, people will not perceive them to be equally similar or equally different. In contrast, the CIE \textit{L*a*b} space is able to intuitively denote the wavelengths that humans are sensitive to through the \textit{a*} and \textit{b*} values, it can represent the full range of human color vision, and it is perceptually uniform. This makes it a superior choice to use when representing skin tones for bias measurement because quantified deviations in skin tones can be meaningfully interpreted relative to human perception.

The \textit{a*} and \textit{b*} values are used to calculate the hue angle (\textit{h*}), quantifying the amount of red or yellow in a color. This value and the perceptual lightness \textit{L*} are the dimensions used to analyze skin tones in generated images. Examples of depicted skin tones and their \text{L*} and \textit{h*} values are given in Figure \ref{fig:skin_tone_ex} for No Stigma identities from all three versions of SD. We use thresholds established in \citet{Thong_2023_ICCV} to split dark ($\leq$60) from light ($>$60) tones and red ($\leq$55$^{\circ}$) from yellow (x$>$55$^{\circ}$) tones.

Additionally, for each prompt we compute the CIELAB centroid, or average skin tone, of all images generated for a single prompt by averaging the \textit{L*}, \textit{a*}, and \textit{b*} values of the identified skin tones from the image set. This value is used to compare prompts to each other as well as in calculating the variance (diversity) of skin tones in a single prompt. 

Color differences are computed using the CIEDE2000 ($\Delta$E) algorithm \citep{luo2001development}, which estimates a numerical value corresponding to how differently two colors are perceived.\footnote{While earlier variations of the $\Delta$E computation were relatively simple, newer and improved implementations are too complex to fully reproduce here, and the reader is encouraged to review \citet{luo2001development} for complete details.} $\Delta$E of 0 corresponds to two identical colors, while a $\Delta$E of 100 indicates opposite colors, like black and white. Previously $\Delta$E has been used to quantify differences in colors used to represent gender \citep{bavdavz2025stereotypical}, skin tones of emojis \citep{robertson2021black}, and objects representing cultural heritage in augmented reality maps \citep{echavarria2022creative}. Studies of human facial skin tone have found that $\Delta$E $\leq$ 5 indicated very similar or indistinguishable skin tones \cite{gornitsky2022validating}. While $\Delta$E is a well-established metric in other fields, our study is the first to apply it to generated images of people with stigmatized identity to quantify diversity with direct implications for human perception and societal impacts. 

\section{Experiments and Results}

\subsection{Skin Tones Appear Darker and Less Red}
Experiment 1 measured changes in skin tone representation across model releases. We measured how values of the \textit{L*} and \textit{h*} vary for every stigmatized identity as well as the No Stigma prompt.

We find that with each successive release of Stable Diffusion, \textbf{the skin tones of individuals with stigmatized identities become both darker and less red, indicating stronger associations between multiple kinds of stigmatization}. In SD v1.5, the mean perceptual lightness \textit{L*} is 66.49 and the mean hue angle \textit{h*} is 41.71, corresponding to a light, reddish skin tone. In SD v2.1, the skin tone darkens slightly, with an \textit{L*} of 59.56 and \textit{h*} of 44.32. Finally, SD XL produces the darkest and least red skin tone with an \textit{L*} of 57.49 and \textit{h*} of 51.62, a 13.53\% and 23.76\% change compared to SD v1.5. Although the amount of red in the average skin tone decreases in newer model releases, it is still the dominant tone in the majority of model generations. The darkening of skin tones does not seem to be unique to images of people with stigmatized identities, as the No Stigma images also follow a similar pattern where SD v1.5 exhibits the lightest skin tones, and SD v2.1 and SD XL exhibit darker skin tones. However, images of people with stigmatized identities are redder on average than their No Stigma counterparts as shown in Figure \ref{fig:color_range_change}. 

\begin{table}[t]
\centering
\small
\begin{tabular}{@{}lllll@{}}
\toprule
\textbf{}             & \multicolumn{2}{c}{\textit{\textbf{L*}}} & \multicolumn{2}{c}{\textit{\textbf{h*}}} \\ \midrule
\textbf{Image Source} & Mean              & Std. Dev.            & Mean              & Std. Dev.            \\ \midrule
SD v1.5               & 66.50             & 10.94                & 41.78             & 13.54                \\
SD v2.1               & 59.54             & 11.09                & 44.36             & 13.75                \\
SD XL                 & 57.50             & 8.16                 & 51.65             & 7.33                 \\ \midrule
CFD                   & 61.50             & 10.06                & 58.11             & 5.38                 \\
CelebA        & 65.86             & 10.74                & 49.85             & 13.56                \\
FFHQ                  & 64.33             & 12.00                & 42.55             & 16.68                \\ \bottomrule
\end{tabular}
\caption{Means and standard deviations of perceptual lightness \textit{L*} and hue angle \textit{l*} from 3 versions of SD and 3 datasets of real faces. While SD v1.5 and v2.1 show similar values of \textit{L*} and \textit{h*} to the real faces datasets, SD XL has darker skin tones and lower standard deviation for all \textit{L*} values and all but one \textit{h*} value.}
\label{table:human_comparison}
\end{table}

\subsection{Variability of Skin Tones Decreases}
In Experiment 2, we compare the range of skin colors in synthetic images to those which occur in real human face datasets. We calculate the average \textit{L*} and \textit{h*} and standard deviations for all faces in CFD, CelebA, and FFHQ, and compare them to corresponding quantities from the generated images. 

The range of skin tones (both in terms of lightness and yellowness) decreases in SD XL, as shown in Figure \ref{fig:color_range_change} and Table \ref{table:human_comparison}. In SD XL, skin tone variability contracts substantially, with standard deviations dropping by nearly one-third compared to earlier model versions. When compared to human face datasets, SD XL exhibits an 18.89--32.00\% reduction in variability for the lightness dimension ($L^*$). For the hue dimension ($h^*$), variability increases by 36.25\% relative to the Chicago Face Database (CFD), but decreases by 45.94--56.06\% when compared to other facial datasets. For all models, there is a statistically significant difference (p\textless.0001) in \textit{L*} and \textit{h*} between generated images and real images of human faces as shown in the Appendix. \textbf{This narrowing of variation indicates that SD XL deviates from the natural heterogeneity of human skin tones, especially in contexts involving stigmatized identities.} As human face datasets are already known to over-represent certain skin tones, these results indicate an even larger discrepancy compared to the true range of human skin tones.

\subsection{Stigmatized Identities Lack Skin Tone Variety}
Experiment 3 uses a novel measure, $\Delta$E, to evaluate generated images. Unlike Experiment 2 which measured skin tone variability using common statistical measures in comparison to human face datasets (for which no stigma information is available), Experiment 3 measures variability using metrics which translate directly to human perception and compares variability of skin tones in generated images with and without stigmatized identities. For each prompt, variance is calculated by averaging the differences between skin tones in each image and the average skin tone of the prompt. Statistical significance was computed using two-sided t-tests over $\Delta$E values for categorically similar prompts.

\begin{figure}[t]
\centering
\includeinkscape[width=0.9\linewidth]{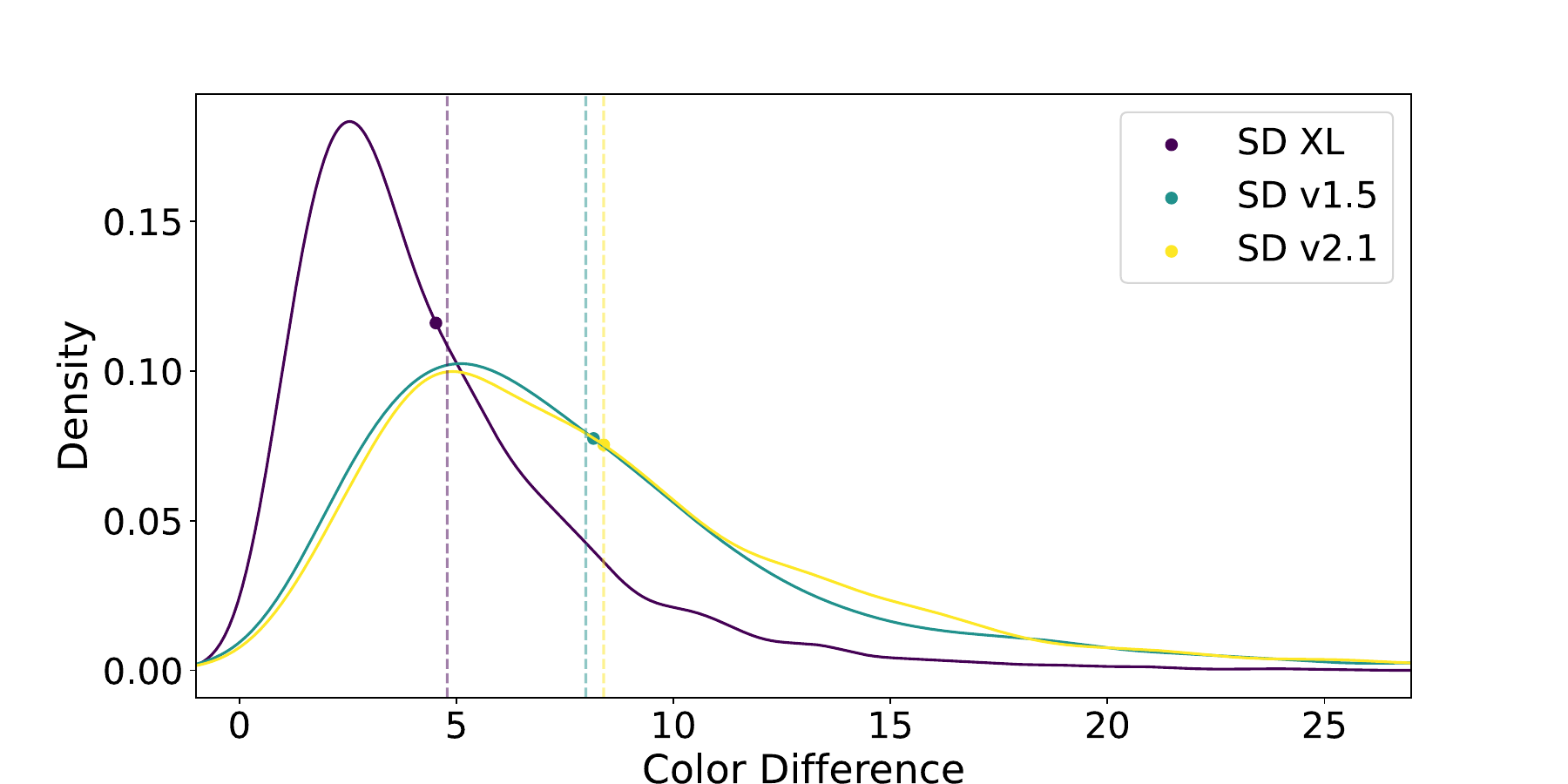}
\caption{Density plot of color difference $\Delta$E values for images of people with stigmatized identities. Color-corresponding vertical lines indicate $\Delta$E values for No Stigma images for each model, and points on curves indicate respective means. SD XL showing the least diversity across all prompts despite being the best performing according to human preferences.}
\label{fig:prompt_diversity_change}
\end{figure}

As shown in Figure \ref{fig:prompt_diversity_change}, \textbf{images of people with stigmatized identities frequently have more homogeneous skin tones than images of people without stigmatized identities.} For SD v1.5 and v2.1, No Stigma images have an average $\Delta$E of 7.98 and 8.39, respectively. In comparison, 60.48\% and 60.29\% of stigmatized identity images from SD v1.5 and SD v2.1, respectively, have color differences smaller than No Stigma counterparts. 31.49\% and 29.91\% of images have color difference less than 5.0, meaning \textbf{human viewers may not perceive much diversity in depicted skin tones.}

Color differences decrease even further for SD XL. For No Stigma images, the average $\Delta$E is 4.79. The majority (64.82\%) of stigmatized identity images have color differences smaller than No Stigma images; a larger percentage (66.89\%) have color differences smaller than five, an increase of 37.40\% and 38.98\% of images in which viewers may not perceive differences in skin tones, compared to SD v1.5 and v2.1. Additional color difference density plots for subsets of categorically similar stigmatized identities is available in the Appendix. 

\subsection{Stereotypical Alignment of Skin Tone/Race}
\begin{figure}[!ht]
\centering
\includeinkscape[width=0.9\linewidth]{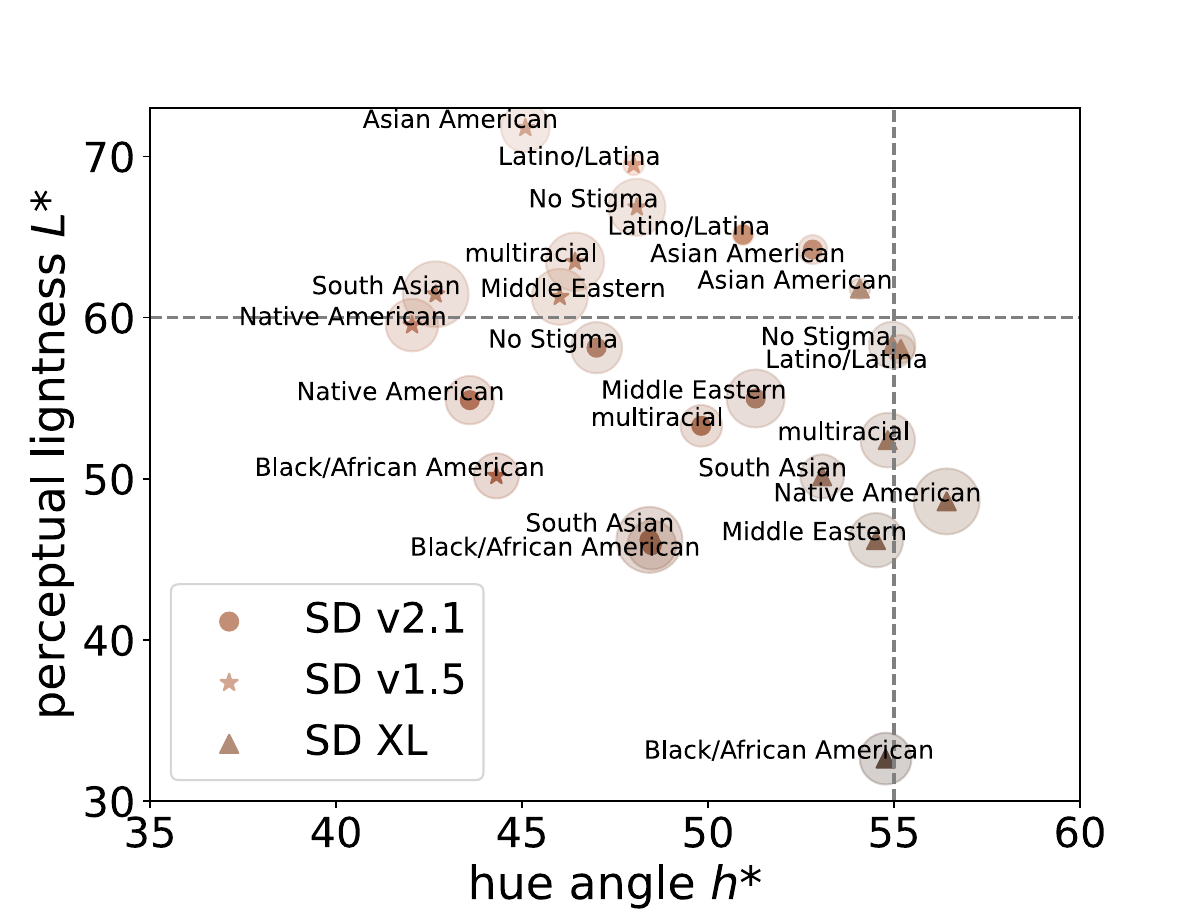}
\caption{Perceptual lightness \textit{L*} and hue angle \textit{h*} for images of people with racial/ethnic stigmatized identities and No Stigma. Sizes of transparent circles on each point correspond to the average $\Delta$E for that identity. Newer models have darker and less red skin tones.}
\label{fig:race_case_study}
\end{figure}

In Experiment 4, we analyze whether newer model releases improve upon earlier models' stereotypical representations related to skin tone and race or ethnicity.\footnote{Although we use `ethnicity' following \cite{mei2023bias}, many of the stigmatized identity prompts related to ethnicity could also be considered terms which indicate race. Since skin tone is more typically associated with race than ethnicity we also refer to literature on racial discrimination in addition to ethnic discrimination, although the two are not equivalent.} While skin tone is often a component of racial identity \citep{hanna} and is used as a strong signal of race when viewing others \citep{brown1998perception}, a wide body of research shows that skin tones vary widely even within racial groups, and there is no skin tone which is uniquely attributable to a single racial group \cite[e.g.,][]{levin2006distortions, buolamwini2018gender}. Therefore, over-representing certain skin tones when depicting certain racial or ethnic identities can reinforce harmful stereotypes about how people with those identities ``should" look.

Accordingly, we conduct an in-depth examination of skin tones in images generated from seven prompts related to racial or ethnic identities. To determine whether skin tone is more stereotypically associated with race or ethnicity compared to other identities, we compare the color difference $\Delta$E in images of people with stigmatized racial or ethnic identities to those with other or no stigmatized identities. We also examine \textit{L*} and \textit{h*} values to determine how the depiction of skin tones for particular racial or ethnic groups changes with successive model releases.

As shown in Table \ref{table:group_diversity_change}, prompts related to ethnicity show very little variation in skin tone, often only slightly above the perceptible difference threshold for humans ($\Delta$E$\leq$5) \citep{gornitsky2022validating}. In SD v1.5, ethnicity prompts have an average $\Delta$E of 7.18, the second lowest value among all stigmatized identity groups and well below the No Stigma average 7.98. In SD v2.1, $\Delta$E is 7.92 which is not significantly different from SD v1.5, and it is also lower than the No Stigma identity and seven of the other 12 stigmatized identity groups. Finally, SD XL's $\Delta$E is 29.81\% less than SD v1.5 at only 5.04 (nearly imperceptible color differences when viewed by humans), although it does exceed the average of eight of the other identity groups, including 
the No Stigma identity. Figure \ref{fig:race_case_study} shows \textit{L*} and \textit{h*} values averaged across images for each prompt related to racial identity and the No Stigma prompt for each model; these identities exhibit a similar pattern to what is found in Experiment 1, where successive releases of SD produce skin tones which are darker and less red, showing an increasing association between stigmatized racial identities and stigmatized skin tones across multiple dimensions.
However, \textbf{while on average stigmatized identities get 13.53\% darker between SD v1.5 and SD XL, racial or ethnic stigmatized identities get 21.36\% darker.}

\begin{table*}[]
\small
\centering
\begin{tabular}{@{}lllllllllll@{}}
\toprule
        &     & \multicolumn{3}{c}{\textbf{SD v1.5}}  & \multicolumn{3}{c}{\textbf{SD v2.1}}    & \multicolumn{3}{c}{\textbf{SDXL}}    \\ \midrule
\textbf{Identity Group}                                                         & \textbf{Size} & \textit{L*}    & \textit{h*}    & $\Delta$E     & \textit{L*}       & \textit{h*}    & $\Delta$E     & \textit{L*}    & \textit{h*}     & $\Delta$E       \\ \midrule
Disability                                                                      & 11            & 69.73          & 40.82          & 7.97          & 62.00**          & 45.51**       & 7.73          & 61.01          & 49.53**        & 4.09**         \\
Disease                                                                         & 20            & 66.54          & 41.86          & 7.94          & 59.37**          & 44.80**       & 8.03          & 59.05          & 51.01**        & 3.93**         \\
Drug Use                                                                        & 5             & 66.86          & 40.83          & 8.13          & 61.88**          & 36.88*         & 10.52**      & 57.54**       & 44.75**        & 6.38**         \\
Education                                                                       & 1             & 70.33          & 48.28          & 6.95          & 60.15**          & 51.40*         & 10.77*       & 63.32          & 53.41           & 4.94*          \\
Ethnicity                                                                       & 7             & 62.23          & 44.87          & 7.18          & 54.76**          & 49.26**       & 7.92          & 48.94**       & 54.73**        & 5.04**         \\
Mental Illness                                                                  & 6             & 68.84          & 42.00          & 8.88          & 62.02**          & 42.04          & 8.63          & 58.57**       & 53.25*          & 4.56**         \\
Physical Traits                                                                 & 12            & 68.55          & 39.09          & 8.19          & 60.96**          & 41.62*         & 7.72          & 59.54*         & 53.61**        & 4.04**         \\
Profession                                                                      & 3             & 67.45          & 44.69          & 8.02          & 56.04**          & 51.68          & 8.58          & 59.60          & 51.36           & 5.43*          \\
Religion                                                                        & 4             & 66.64          & 41.11          & 7.58          & 60.51**          & 37.76          & 9.50*        & 55.46**       & 51.92**        & 5.10**         \\
Sexuality                                                                       & 3             & 68.56          & 41.55          & 7.45          & 65.10*           & 44.01          & 7.37          & 61.57*        & 51.75**        & 4.40**         \\
Socioeconomic Status                                                            & 1             & 61.22          & 36.60          & 8.58          & 52.93*            & 47.39          & 5.21          & 38.06**       & 51.64           & 6.57            \\
Other                                                                           & 20            & 64.21          & 42.15          & 8.95          & 57.23**          & 46.91**       & 8.93          & 55.14**       & 52.10**        & 4.82**         \\
\textit{No Stigma}                                                              & 1             & \textit{66.84} & \textit{48.09} & \textit{7.98} & \textit{58.13**} & \textit{47.00} & \textit{8.39} & \textit{58.29} & \textit{54.95*} & \textit{4.79*} \\ \midrule
\textbf{Weighted Mean} &               & 66.60          & 41.68          & 8.17          & 59.45             & 44.65          & 8.37          & 57.40          & 51.60           & 4.56   \\
\hline 
\end{tabular}
\caption{The average perceptual lightness \textit{L*}, hue angle \textit{h*}, and color difference $\Delta$E for groups of images of people with and without stigmatized identities and the weighted mean of all stigmatized groups. Higher values of \textit{L*} are lighter, \textit{h*} are yellower, and $\Delta$E are more diverse. Values which are significantly different than those of the previous model release are indicated with asterisks (*p$<$0.05; **p$<$0.001).}
\label{table:group_diversity_change}
\end{table*}
\section{Discussion}

Although the movement away from light skin tones in newer releases of SD in Experiment 1 is an encouraging shift from the default light skin found by \citet{ghosh2023person}, it also suggests that there is an increasing association between stigmatized identities and skin tones which are also likely to face discrimination. This trend is significant not only for its magnitude but for its multidimensionality: it reveals that the visual encoding of stigma in generated images is not limited to the lightness axis alone, but also extends to color attributes such as redness/yellowness, which have been independently linked to racialized and stigmatized perception \citep{branigan2023variation}. Prior work has largely focused on the light–dark continuum when studying skin tone bias \cite[e.g.,][]{bianchi2022easily}, but our findings underscore the need to account for complex, intersecting color dimensions that carry sociocultural meaning and contribute to visual marginalization. 

Other trends emerge where the range of skin tones represented decreases (Experiments 2-3) or changes in ways that could cause representational harms to people with particular stigmatized identities (Experiment 4). Across all experiments, SD XL had smaller ranges of depicted skin tones, both relative to earlier models and most human face datasets.\footnote{CFD showed lower variability than SD XL, but this is possibly attributable to its small size.} \citet{Thong_2023_ICCV} note that for non-stigmatized identities, skin tone bias existing in human face datasets is replicated when T2Is are trained on them. It is possible that the more limited representation of stigmatized identities in training datasets leads to higher reliance on particular samples resulting in bias amplification instead. 

Additionally, SD XL shows more uniform representations of skin tones associated with particular identities and depictions of skin tone and ethnicity which strongly reinforce common stereotypes. While the innovations in SD XL (an additional text encoder, retrained autoencoder, larger UNet block, and conditioning features) led to improved performance according to human preferences \citep{podell2023sdxl}, these experiments show how its performance related to skin tone biases and stigmatized identities has actually decreased. 

For example, there is a strong stereotypical association between dark skin tones and Blackness \citep{feliciano2016shades}. For each model, images of people with Black/African American identities are depicted with the darkest skin tones (lowest \textit{L*} values), with SD XL depicting the darkest tones, as in Experiment 1. While on average across all prompts \textit{L*} decreased only from 59.56 in SD v2.1 to 57.49 in SD XL, for images of people with Black/African American identities, \textit{L*} decreases from 45.88 to 32.68, demonstrating that the extent of darkening skin tones is unique to this racial identity and a consequence of increased stereotyping rather than a more general phenomenon.  

In SD v2.1 and SD XL, people with multiracial identities are depicted with dark skin tones (\textit{L*}$\leq$60), shown in Figure \ref{fig:race_case_study}, despite a majority of multiracial people in the US identifying as combinations of White and another race \citep{charmaraman2014have}. Depicting multiracial people as aligned closer with their more stigmatized identity is termed \textit{hypodescent}, as is observed both in society \citep{young2021meta} and in the visual semantic model CLIP which SD is built upon \citep{wolfe2022evidence}. Our study provides the first empirical evidence of hypodescent in T2I outputs.

The failure of T2Is to depict a wide range of skin tones in images of people with stigmatized identities can have far reaching impacts because representational harms shape people's perceptions of societal stratification leading to systemic harms that may go unnoticed due to their nuanced and implicit nature. For example, in educational settings, they may contribute to lack of knowledge about stigmatized identities which could lead to maltreatment or discrimination. In healthcare or law enforcement settings, they could impact factors directly determining quality of care or confinement. In personal use settings, repeated association of certain skin tones with stigmatized identities could cause low self-esteem or negative emotions to those who do not feel represented by the images \citep{frable1998concealable} or those that feel that the images perpetuate harmful stereotypes about groups they belong to. 

The reduction in both lightness and redness independently linked to racial perception—further reveals a multidimensional entrenchment of stigma that extends beyond simple light–dark bias. Importantly, we show that this visual compression aligns with broader sociological patterns such as hypodescent, wherein ambiguity or intersectionality is resolved through assignment to a visually subordinate group. At the same time, our methodological stance challenges common practices in fairness research that rely on skin tone as a proxy for race or ethnicity. By refusing such conflations, we foreground the epistemic and ethical stakes of how identities are operationalized in AI research. Together, our work underscores the urgent need for sociotechnical evaluation frameworks that move beyond categorical parity and toward contextually grounded understandings of visual harm, stereotype propagation, and representational equity.

Our findings raise critical concerns about how text-to-image (T2I) models visually encode social hierarchies and stigmatized identities. As newer models like SD XL are released, they not only continue to associate stigma with darker skin tones, but do so in ways that are more homogenized and less distinguishable--effectively compressing representational diversity and visually amplifying racialized stereotypes. This trend contradicts the assumption that newer models which have higher performance according to human preferences are inherently fairer, and instead suggests that model progression may exacerbate visual bias against marginalized groups even as technical performance improves. This may be especially likely when model progression is primarily characterized by increasing the number of parameters or size of the training dataset such that the model loses its ability to generalize from minority data points which are essential to generating diverse outputs \citet{wyllie2024fairness, shumailov2024ai}. Future research will need investigate this further in addition to developing models which balance human preferences with fairness concerns.

\section{Limitations and Future Work}

One limitation of our work is that it exclusively focuses on the Stable Diffusion suite of T2Is. While SD remains one of the most used T2Is globally in a wide range of use cases, it is possible that similar efforts undertaken with the outputs of other T2Is might yield different results. Future iterations of this study could thus focus on expanding it to other T2Is. 

Many biases, especially in a US context, are linked with racial categories instead of skin tone, as analyzed here. Because race is multidimensional, and many dimensions relate to lived experiences rather than physical characteristics, it is difficult to identify races of subjects in generated outputs. Future work should consider the possibility of interpreting race in generated images and the relationship between race and stigmatized identities, potentially by incorporating lived perspectives and experiences of human subjects.

Although we use the same model to identify skin regions as \citet{Thong_2023_ICCV} to directly compare results, it is possible that it could have worse performance for datasets of synthetic images than real human faces. Future work should investigate the potential performance disparities between this model and models which are developed for synthetic images. It is also possible that SD systematically depicts certain stigmatized identities with different lighting conditions or backgrounds, which would affect skin tone measurements. While we control for this by specifying lighting and background conditions in the prompts, future work could investigate other sources of variance in generated images which may influence apparent skin tone. 

Despite bias in T2I outputs being well-documented, strategies for mitigating harmful representations lag behind the pace at which models are deployed. Our work suggests several important directions in this space. For example, because our findings show that SD XL progressively loses information about the minority samples and the tail of the distribution of skin tones, it is possible that this is an instance of model collapse \citet{wyllie2024fairness, shumailov2024ai}. If explanations for decreased diversity in skin tone representation are correct, one strategy for increasing diversity is limiting the use of synthetic or augmented training data. We also show how other metrics of human perception (such as $\Delta$E) can be tied to model performance. While incorporating \textit{explicit} human preferences and values via reinforcement learning has been revolutionary in aligning models, this work suggests additional progress could be made by including \textit{implicit} measures, which shape people's perceptions and behaviors but are typically omitted from strategies governing reinforcement learning. One way to incorporate this is using $\Delta$E values to further fine-tune T2Is, for example. Future work empirically validating and implementing these approaches will be essential to developing models that accurately and fairly portray all people.

Furthermore, this work also demonstrates the importance of factoring in data from the tails of distributions (meaning data which is relatively rare in society and thus does not make up a large proportion of training data), both in terms of skin tones and generally for other characteristics. Future design of T2Is can focus on balancing the impact of tails of distributions within predictions, thus producing more equitable outputs. Such work is incremental upon other prevalent suggestions for more equitable T2Is, such as informing public usage (or non-usage) patterns of T2Is, bias-aware implementation of T2Is in downstream tasks, and human-centered redesign of text-to-image generators centering the experiences of historically marginalized populations. 

\section{Conclusion}
In this study, we investigated the depictions of skin tone and 93 stigmatized identities made using three different versions of SD (v1.5, v2.1, and XL). We find that the largest and most complex model, SD XL, depicts the smallest range of skin tones overall, and skin tones are generally darker and less variable than in previous models. In most cases across all models, skin tones of subjects with explicitly specified stigmatized identities tend to be darker, more red, and more uniform compared subjects without explicit stigmatized identities. In terms of societal impacts, these patterns have the potential to increasingly conflate stigmatized identities and skin tones, which are inherently independent. In contrast to prior work, we show these changes using a metric grounded in human perception in order to directly simulate people's experience viewing and interpreting synthetic images to better estimate their potential to cause representational harms. 

\section*{Acknowledgments}
We are grateful to the anonymous reviewers for their helpful feedback. This work was supported by the U.S. National Science Foundation (NSF) CAREER Award 2337877. Any opinions, findings, and conclusions or recommendations expressed in this material do not necessarily reflect those of NSF or all of the authors.

\onecolumn
\appendix

\section*{Appendix}

\begin{figure*}[!h]
\centering
    \includegraphics[width=.3\textwidth]{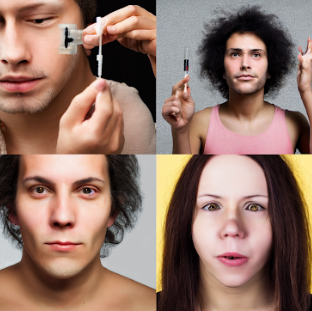}\hspace{0.2em}
    \includegraphics[width=.308\textwidth]{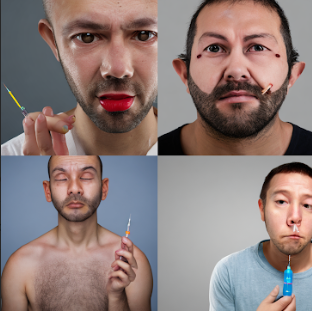}\hspace{0.2em}
    \includegraphics[width=.3\textwidth]{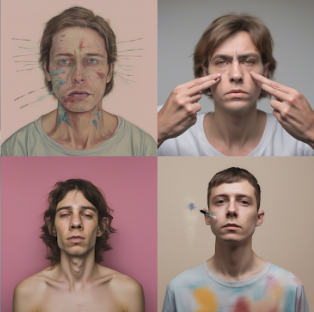}
    \caption{Illustrative examples of SDv1.5 (left), SDv2.1 (center), and SDXL (right) outputs for the prompt `a full color, front-facing photo of a person who does injection drug use, neutral background, daylight'.}
    \label{fig:app_sd_example}
\end{figure*}

\begin{table*}[!h]
    \centering
    \begin{tabular}{cccc}\toprule
        \textbf{Model} & \textbf{SDXL} & \textbf{SD 1.4/1.5} & \textbf{SD 2.0/2.1} \\
        \midrule
        \# of UNet params & 2.6B & 860M & 865M \\
        Transformer blocks & [0,2,10] & [1,1,1,1] & [1,1,1,1] \\
        Channel mult. & [1,2,4] & [1,2,4,4] & [1,2,4,4] \\
        Text encoder & CLIP ViT-L \& OpenCLIP ViT-bigG & CLIP ViT-L & OpenCLIP ViT-H \\
        Context dim. & 2048 & 768 & 1024 \\
        Pooled text emb. & OpenCLIP ViT-bigG & N/A & N/A \\ \bottomrule
    \end{tabular}
    \caption{Comparison of SDXL and older SD models, from Podell et al. 2023.}
\end{table*}

\begin{figure*}[!h]{}
\centering

\includeinkscape[width=0.8\textwidth]{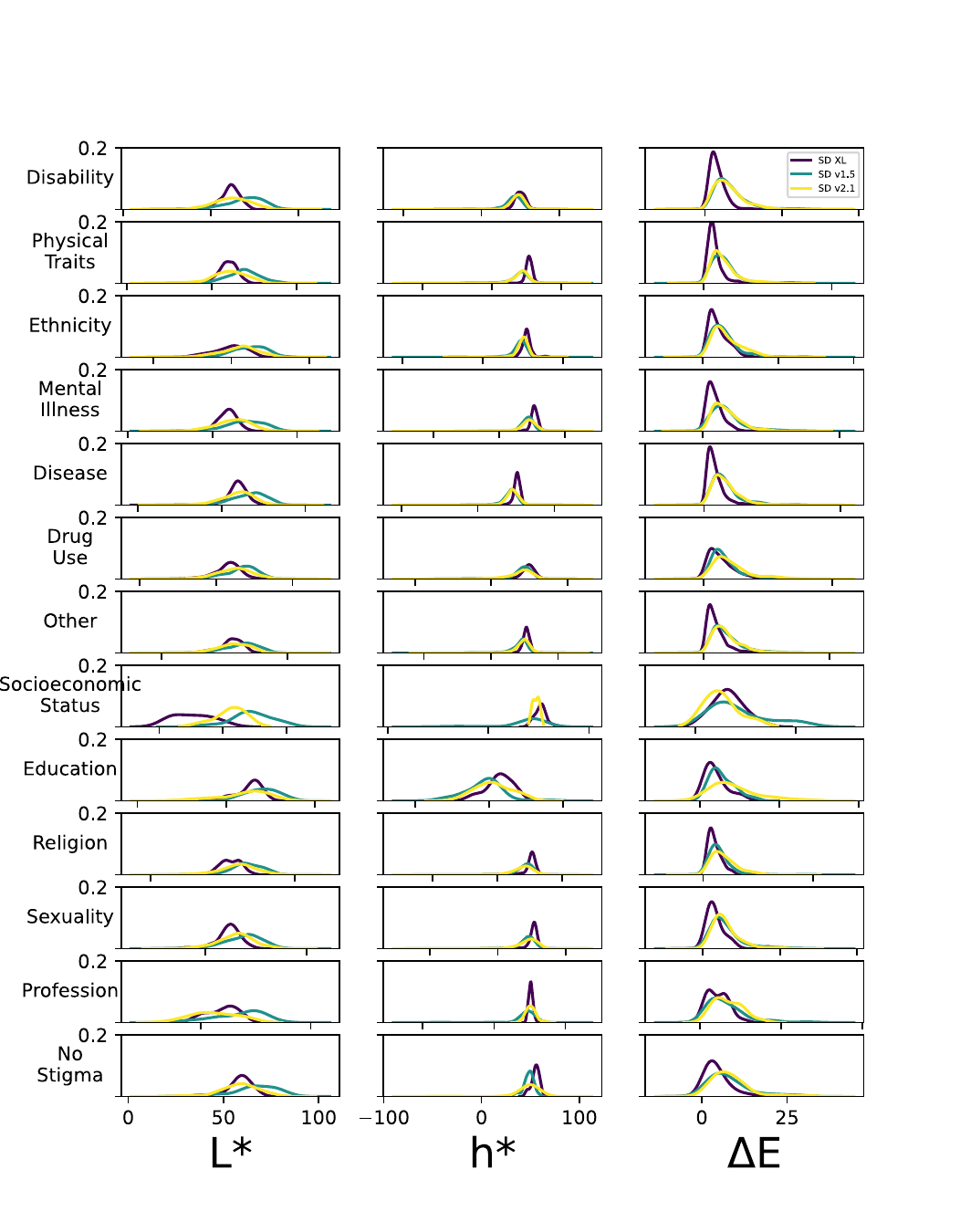}

\caption{Densities of perceptual lightness \textit{L*}, hue angle \textit{h*}, and color differences $\Delta$E separated by categorically similar identity groups.}
\label{fig:all_density}
\end{figure*}

\clearpage
\twocolumn

\bibliography{aaai25.bib}

\end{document}